\documentclass[aps, prl, showpacs, floatfix, preprint]{revtex4-1}
\usepackage{graphicx,color,hyperref}
\begin{document}

\title{Sensitive BEC magnetometry with optimized probing}
\author{Naota Sekiguchi}
\affiliation{Department of Physics, 
Gakushuin University, Tokyo, Japan}
\author{Aki Torii}
\affiliation{Department of Physics, 
Gakushuin University, Tokyo, Japan}
\author{Hiroyuki Toda}
\affiliation{Department of Physics, 
Gakushuin University, Tokyo, Japan}
\author{Ryohei Kuramoto}
\affiliation{Department of Physics, 
Gakushuin University, Tokyo, Japan}
\author{Daiki Fukuda}
\affiliation{Department of Physics, 
Gakushuin University, Tokyo, Japan}
\author{Takuya Hirano}
\affiliation{Department of Physics, 
Gakushuin University, Tokyo, Japan}
\author{Kosuke Shibata}
\affiliation{Department of Physics, 
Gakushuin University, Tokyo, Japan}
\email{shibata@qo.phys.gakushuin.ac.jp}

\date{\today}

\begin{abstract}
An improved spatial magnetometer using a spinor Bose--Einstein condensate of $^{87}$Rb atoms is realized utilizing newly developed two-polarization phase contrast imaging.
The optical shot noise is suppressed by carefully choosing the probe parameters.
We attain a dc-magnetic field sensitivity of $7.7~\mathrm{pT/\sqrt{Hz}}$ over a measurement area of $28~\mathrm{\mu m^{2}}$.
The attained sensitivity per unit area is superior to that for other modern low-frequency magnetometers with micrometer-order spatial resolution.
This result is a promising step for realizing quantum-enhanced magnetometry surpassing classical methods.
\end{abstract}
\pacs{}

\maketitle

High-performance magnetometry is of great importance in fundamental research and applications. 
While sensitivities of sub $\mathrm{fT/\sqrt{Hz}}$ have been achieved using optically pumped magnetometers \cite{Kominis2003,Dang2010} and superconducting quantum interference devices (SQUIDs) \cite{Schmelz2011} with spatial resolutions of several to tens of millimeters, nanometer spatial resolutions have been attained using single trapped ions with sensitivities of a few $\mathrm{pT/\sqrt{Hz}}$ \cite{Baumgart2016} and single nitrogen-vacancy centers in diamond with several $\mathrm{nT/\sqrt{Hz}}$ sensitivity \cite{Balasubramanian2009}.

A cold atom cloud is also a good medium for spatial magnetometry.
In particular, the Bose--Einstein condensate (BEC) is a promising candidate for pursuing high-performance spatial magnetometry because of its ultimately controlled position and momentum uncertainty. A thermal gas magnetometer of micrometer resolution using a shaped optical potential has been demonstrated \cite{Eliasson2019}, and BEC magnetometers with micrometer resolutions and picotesla sensitivities have been realized \cite{Wildermuth2005, Vengalattore2007}.
A spinor BEC with a long spin coherence time of a few seconds \cite{Jasperse2017,Palacios2018} is attractive for sensitive low-frequency field detection, which is important for several applications such as biomagnetic diagnostics \cite{Hamalainen1993} and measurements of magnetic properties of a solid-state sample \cite{Yang2020,Hu2020}.

Considering the field sensitivity and spatial resolution that can be achieved, a BEC magnetometer is an excellent platform to investigate the limit of magnetic field sensitivity per measurement volume, which has not yet been clarified \cite{Mitchell2020}.
Moreover, a BEC magnetometer is capable of quantum enhanced measurements \cite{Sewell2012,Muessel2014,MartinCiurana2017,Colangelo2017,Chalopin2018,Evrard2019} and can be used to study the relation between the limit of sensitivity  and quantum noise.

The sensitivity of BEC magnetometry has been limited by the shot noise of the probe light \cite{Vengalattore2007}.
Increasing the probe-light intensity improves the signal to noise ratio associated with photon shot noise.
However, a brighter probe light has a greater influence on the atoms in the target, causing optical pumping, nonlinear spin evolution \cite{Smith2004, Deutsch2010, Colangelo2013}, and superradiant Raman scattering \cite{Vengalattore2007}, which result in atom loss and signal decay.
Detailed investigations of the light-induced loss and decay are therefore required for further improvements in the performance of BEC magnetometers.
Recently, the importance of properly selecting the frequency of the near-resonant $D_{1}$ light was  demonstrated in the context of laser cooling \cite{Urvoy2019}.

We conducted a comprehensive survey of the light-induced decay properties of Larmor precession in a BEC of $^{87}$Rb atoms using a newly developed technique of two-polarization phase contrast imaging (TPPCI).
TPPCI enabled us to simultaneously track the atom number and spin density in a single measurement.
We found that the decay was suppressed for a carefully chosen probe frequency at the red detuned side, which suggests that the decay is associated with light-induced collisions.
After probe parameter optimization, we achieved a precession phase sensitivity of 5.6~mrad over a measurement area of 142~$\mu$m$^2$.
The magnetic field sensitivity was also evaluated to be $7.7~\mathrm{pT/\sqrt{Hz}}$ over a measurement area of $28~\mathrm{\mu m^{2}}$.
The phase and field sensitivities are better than those for state-of-the-art spatial BEC magnetometry \cite{Vengalattore2007}.

\begin{figure}[t]
 \centering
\includegraphics[width=8.6cm]{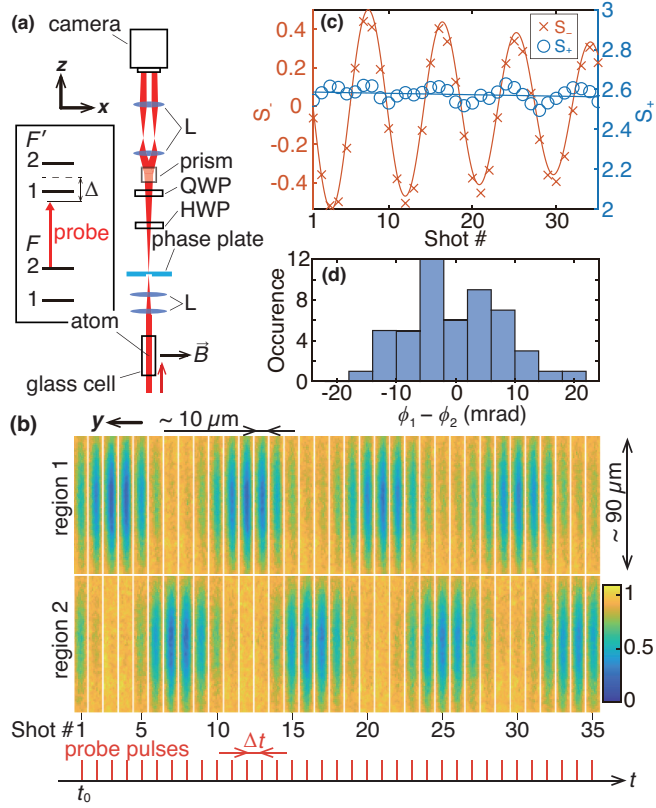}	
 \caption{(a) Illustration of the TPPCI setup. L: lens, HWP: half-waveplate, QWP: quarter-waveplate. The inset shows the energy diagram of $^{87}$Rb and the frequency of the probe light.
 (b) Typical set of contrast images.
The top and bottom rows display the orthogonal polarization components and are sorted by time from left to right.
The color bar represents the contrast.
(c) Difference ($S_{-}$) and sum ($S_{+}$) signals.
The lines are fits to the data.
(d) Histogram of $\phi_{1}-\phi_{2}$ over 50 experimental runs.
}
\label{fig: typ}
\end{figure}

TPPCI was implemented with the setup schematically depicted in Fig.~\ref{fig: typ}(a). 
A BEC of $^{87}$Rb in the hyperfine ground state $F=2$ was trapped in an optical dipole trap in a glass cell.
The number of atoms in the BEC was approximately $3\times 10^{5}$.
The probe light was linearly polarized and propagated along the $z$-axis perpendicular to the bias magnetic field $\vec{B}~(\parallel \hat{x}$) of about $14~\mu$T.
The angle $\alpha$ of the polarization plane with respect to $\vec{B}$ was chosen around the magic angle of 54.7$^{\circ}$ to minimize light-induced nonlinear spin evolution \cite{Smith2004,Deutsch2010,Colangelo2013}.
The probe light frequency was stabilized with a tunable detuning $\Delta$ from the $D_{1}$ line  transition ($\lambda = 795$ nm) of $^{87}$Rb.
$\Delta$ was defined with respect to the frequency from the $F=2$ state to the center of the excited states $F'=1,2$.
The probe-beam $1/e^{2}$ diameter of 3~mm was sufficiently larger than the axial atom-cloud size of $\sim$90~$\mu$m.

A phase-contrast image was formed using a phase plate with a tiny dimple ($97.9~\mathrm{\mu m}$ in diameter) at the center.
The phase-contrast image was decomposed into orthogonal circularly polarized components by waveplates and a Nomarski prism. 
The polarization components were focused on different regions (region 1 and 2) on a CCD camera.

Figure~\ref{fig: typ}(b) shows a typical set of contrast images taken by 35 probe pulses.
The spin state of the BEC was initially polarized along $\vec{B}$ and rotated onto the $y$--$z$ plane by an rf $\pi/2$ pulse.
After a given time $t_{0}$, the BEC was imaged with probe pulses evenly spaced by $\Delta t$.
The probe-pulse width $\tau_{p}$ was less than $1~\mathrm{\mu s}$. 
$\tau_{p}$ was sufficiently shorter than the Larmor period $\tau_\mathrm{L}= 2\pi\hbar/(g\mu_{B}|\vec{B}|) \sim 10~\mathrm{\mu s}$, where $\hbar$ is the reduced Planck constant, $g$ is the Land\'e g-factor, and $\mu_{B}$ is the Bohr magneton.
For Fig.~\ref{fig: typ}(b), we set the parameters as follows: 
$t_{0} = 0.1~\mathrm{ms}$, $\Delta t = 30~\mathrm{\mu s}$, $\tau_{p}=600~\mathrm{ns}$, photon fluence $\Phi_{p}$ per pulse of $8.3\times10^{3}~\mathrm{\mu m^{-2}}$, and $\Delta=-1.537~\mathrm{GHz}$.

The temporal oscillation in the contrast signal in each region indicates the spin precession about the $x$-axis (aliased by the sampling rate $1/\Delta t$).
For a qualitative analysis, we integrated the contrast signals within a region of interest around the center of the BEC.
We then took the sum $S_{+}$ and difference $S_{-}$ of the integrated signals at both regions.
The results are shown in Fig.~\ref{fig: typ}(c).
With the assumption that the phase shifts caused by the atoms are sufficiently smaller than unity, $S_{+}$ and $S_{-}$  are approximated as
\begin{eqnarray}
S_{+} &\simeq& 2+4\theta_{0}+2\theta_{l}\cos2\alpha, \label{eq: sp}\\
S_{-} &\simeq& 2\theta_{c}. \label{eq: sm}
\end{eqnarray}
Here, $\theta_0$ is the common phase shift, $\theta_{c}$ is the phase difference between circular polarization components,  and $\theta_{l}$ is the phase difference between linear polarization components.

Since $\theta_c$ is proportional to the projection of the spin on the probe propagation axis, $S_{-}$ is a direct measure of the Larmor oscillation.
The damped oscillation of $S_-$ shown in Fig.~\ref{fig: typ}(c) indicates that the Larmor signal gradually decreased due to the probe pulses.
The behavior of $S_+$ can be understood from Eq.~(\ref{eq: sp}).
The gradual decrease of $S_{+}$ was attributed to the loss of atoms, which decreases $\theta_0$.
The oscillation in $S_+$ partly results from the spin alignment that leads to oscillation in $\theta_{l}$.
The slight increase of the oscillation amplitude suggests that orientation-to-alignment conversion was induced by the probe light, even though $\alpha$ was set around the magic angle.
The misalignment in the probe system may also be responsible for the oscillation in $S_{+}$.
We note that $S_{+}$ oscillates even with no nonlinear spin evolution due to a term proportional to $\langle F_{z}^{2} \rangle$ in $\theta_{0}$ \cite{Smith2004}.

TPPCI has the following advantages.
First, the spin state and number of atoms can be tracked separately, as discussed, whereas both the spin state and atom number contribute to the signal in other similar imaging methods such as spin-sensitive PCI \cite{Higbie2005, Vengalattore2007} and dual-port Faraday imaging \cite{Kaminski2012}.
Second, $S_{-}$ can be detected without the influence of a probe-intensity fluctuation, due to the balanced configuration. 
Third, the use of a linearly polarized probe eliminates the systematic error due to the vector shift by a circularly polarization component \cite{Vengalattore2007}.
TPPCI can be applied for various purposes including the study of the spatial magnetic properties in a dissipative open system \cite{Eto2019}.
With a different configuration using a half waveplate instead of the quarter waveplate before the prism, it is possible to image the spatial distribution of spin alignment or spin nematicity.
This will be useful for the study of exotic spin states. 

We investigated the atom loss as a function of the detuning $\Delta$ of the probe frequency.
$S_{+}$ was fitted with an exponential function, $A_{+} \exp(-\gamma_{+} t)$, to estimate the atom loss rate $\gamma_{+}$.
We also fitted $S_{-}$ with a damped sinusoidal function, $A_{-}\exp(-\gamma_{-}t)\sin\left(\omega t+\phi_{0} \right)$, where $\gamma_{-}$ is the decay rate of the Larmor precession signal, $\phi_0$ is the initial phase, and $\omega$ is the aliased precession frequency.
The fitting models adopted here are valid while non-exponential decay due to multi-body losses and nonlinear spin evolution are sufficiently small.

\begin{figure}
 \centering
 \includegraphics[width=8.6cm]{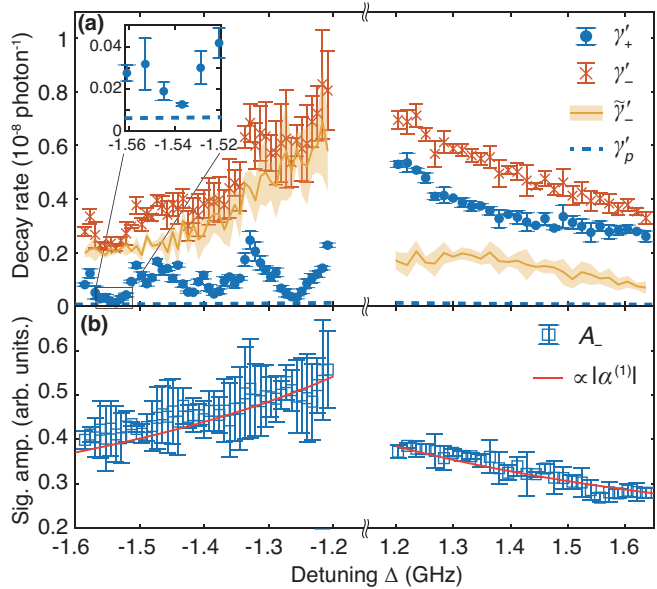}
 \caption{Detuning dependence of the sum and difference signals.
(a) Normalized decay rates. $\gamma_{+}'$, $\gamma_{-}'$, and $\widetilde{\gamma}_{-}'$ are plotted as a function of $\Delta$.
The dotted line represents the photon scattering rate $\gamma_{p}'$.
(b) Larmor signal amplitude.
The solid lines are a fit based on the theoretical frequency dependence.
}
\label{fig: loss}
\end{figure}

The frequency dependences of the decay rates normalized by the number of incident photons, $\gamma_{+}'$ and $\gamma_{-}'$, are shown in Fig.~\ref{fig: loss}(a).
The red- and blue-detuned sides of the frequency dependences were independently measured, and the experimental conditions, such as the environmental magnetic field and the optical trap depth for each side, were slightly different.
The dotted line in Fig.~\ref{fig: loss}(a) shows the normalized photon-scattering rate $\gamma_{p}'$ calculated for our experimental conditions.
It was found that $\gamma_{+}'$ was asymmetric with respect to the sign of $\Delta$. 
The asymmetry in $\gamma_{+}'$ suggests that the atom loss was mainly induced by light-induced collisions \cite{Burnett1996,Fuhrmanek2012,Urvoy2019}, the rate of which is discrete for the red-detuned light and continuous for the blue-detuned light.
The lowest atom loss rate in this work was found to be on the order of $\gamma_{p}'$ around $\Delta = -1.537$~GHz, as shown in the inset.
The peaks of $\gamma_{-}'$ at red-detuned frequencies were associated with the atom loss.

The net decay rate $\widetilde{\gamma}_{-}'$ of the Larmor precession (shown by the solid line with an error band) was estimated to be $\widetilde{\gamma}_{-}'=\gamma_{-}'-\gamma_{+}'$ and showed little discrete dependence on the probe frequency.
The observed net decay rate roughly agreed with a numerical simulation with no superradiance.
This implies that the nonlinear spin evolution is a dominant source of the signal decay in our configuration.

The amplitude $A_{-}$ had a monotonic detuning dependence regardless of the sign of the detuning, as plotted in Fig.~\ref{fig: loss}(b).
The dependence is in good agreement with the theoretical expectation, which is proportional to the absolute value of the vector polarizability $\alpha^{(1)}$ \cite{Geremia2006,Jammi2018} given by 
\begin{equation}
\alpha^{(1)}=-\frac{\epsilon_0 \lambda^3 \Gamma}{64\pi^2} \left[\frac{3}{\Delta +\omega_{\mathrm{hfs}} /2} +\frac{1}{\Delta -\omega_{\mathrm{hfs}} /2} \right],
\end{equation}
where $\Gamma$ is the natural linewidth of the $D_{1}$ line and $\omega_{\mathrm{hfs}} = 2\pi \times 814.5$ MHz is the hyperfine splitting in the excited $P_{1/2}$ states.
Since plausible noise sources such as optical and atomic spin shot noise do not depend on the probe frequency, the signal to noise ratio is maximized at the red-detuned side. 

We evaluated the sensitivity $\delta \phi$ in the Larmor-precession phase with $\Delta = -1.257~\mathrm{GHz}$ and $\Delta = -1.537~\mathrm{GHz}$, which give relatively small $\gamma_{\pm}'$.
To avoid the effect of the common magnetic field fluctuation, including an AC line noise of $< 1~\mathrm{nT}$ at 50~Hz measured by a spin echo method \cite{Eto2013}, $\delta \phi $ was evaluated from $\sqrt{\mathrm{Var}[\phi_{1}-\phi_{2}]/2}$ over 50 experimental runs.
Here, $\phi_{1}$ and $\phi_{2}$ were the estimated initial Larmor phases in two separated regions around the atom cloud center. 
A factor of $2$ was introduced, as $\phi_1$ and $\phi_2$ were expected to be independent.
We fixed the first-pulse time to be $t_{0} = 0.1~\mathrm{ms}$.

A best value of $\delta\phi = 5.6~\mathrm{mrad}$ over an analyzing region of $142~\mathrm{\mu m^{2}}$ was attained for $\Delta = -1.537~\mathrm{GHz}$ and $\Phi_{p} = 8.3\times10^{3}~\mathrm{\mu m^{-2}}$.
The corresponding histogram is shown in Fig.~\ref{fig: typ}(d).
This phase sensitivity was better than the best value for a spatial BEC magnetometer (10~mrad over $120~\mu$m$^2$)\cite{Vengalattore2007}.
Furthermore, this sensitivity was almost at the standard quantum limit $\delta \phi_{\mathrm{SQL}} = \sqrt{\delta \phi_a^2 + \delta \phi_p^2 }\approx 6~\mathrm{mrad}$, where $ \phi_a$ and $\phi_p$ are the phase uncertainties due to the atomic spin and optical shot noises, respectively:
$\delta \phi_a$ was estimated to be $3~\mathrm{mrad}$ in each measurement region and
$\delta \phi_p$ was found to be $5~\mathrm{mrad}$ using independent calibration of the shot noise at the camera and a numerical calculation of the error propagation. 

We also evaluated the performance as a magnetometer.
The magnetic field sensitivity is given by
\begin{equation}
\delta B = \frac{\hbar}{g \mu_B}\frac{\sqrt{T_{\mathrm{cycle}}}}{t_{0}}\delta\phi,
\end{equation}
where $T_{\mathrm{cycle}}$ is the cycle time of a single measurement run.
$T_{\mathrm{cycle}}$ was fixed to 60 seconds in our experiment, regardless of the value of $t_{0}$.
$T_{\mathrm{cycle}}$ includes the time taken for loading atoms into a magneto-optical trap, for the evaporative cooling to a BEC, and for the measurement.

In contrast to conventional spinor BEC magnetometries \cite{Vengalattore2007,Jasperse2017}, we used the BEC in the upper hyperfine state $F=2$.
Although inelastic collisions between atoms in different magnetic sublevels is a concern, it is expected that the polarized spin state does not suffer from inelastic collisional losses.
In fact, the lifetime of the BEC prepared in the transversally polarized state with the first $\pi/2$ pulse was measured to be $T = 0.5\pm0.1~\mathrm{s}$ after minimizing the inhomogeneity in the real and fictitious magnetic fields.
The obtained $T$ was sufficiently long for constructing a sensitive magnetometer but shorter than the lifetime of the longitudinally polarized BEC.
We ascribe the excess atom loss to inelastic collisions combined with a nonlinear spin evolution due to the quadratic Zeeman energy $\approx h\times 2~\mathrm{Hz}$.

\begin{figure}
 \centering
 \includegraphics[width=8.6cm]{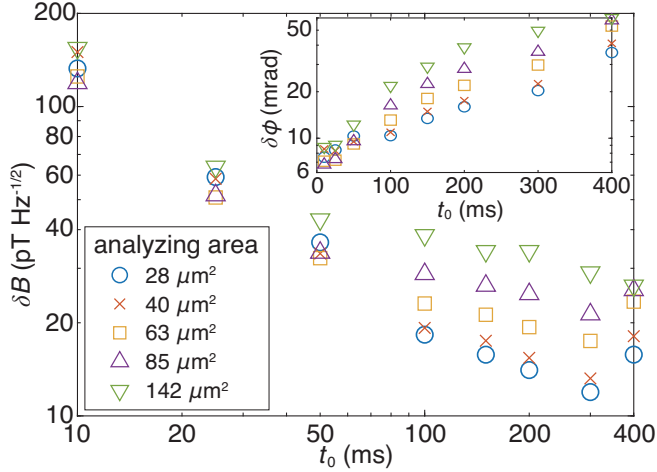}
 \caption{Magnetic field sensitivity $\delta B$ as a function of $t_{0}$ for different analyzing regions.
The inset shows the dependence of the phase sensitivity $\delta \phi$ on $t_{0}$.
}
\label{fig: sensitivity}
\end{figure}

We plotted $\delta B$ and $\delta \phi$ for $\Delta =  -1.537$~GHz and $\Phi_{p}=1.1\times 10^{4}~\mathrm{\mu m}^{-2}$ as a function of $t_0$ in Fig.~\ref{fig: sensitivity}.
The sensitivities were evaluated for different analyzing regions, which had the same height of $5.3~\mathrm{\mu m}$ along the $y$ direction and different widths along the $x$ direction.
We found that a $t_0$ around $300$ ms was optimal for attaining a good field sensitivity
because of the increase in $\delta \phi$ for longer $t_{0}$.
The optimal time was close to the $1/e$-decay time $T_{-}$ of the signal amplitude $A_{-}$, which was measured to be $T_{-}=0.57\pm 0.03~\mathrm{s}$.
We note that $T_-$ monotonically decreased, in contrast to the case of the spinor BEC in the lower hyperfine state, where signal modulation occurs due to a quadratic Zeeman shift \cite{Jasperse2017}.
The monotonic decay should be related to the inelastic collisions \cite{Eto2019}.
The sensitivity was not improved by expanding the analyzing region, against naive expectations.
The sensitivity degraded for a larger region when $t_0 \geq 100$ ms.
We ascribed the sensitivity degradation for larger regions to a magnetic field inhomogeneity.

The best magnetic field sensitivity of $\delta B=7.7~\mathrm{pT/\sqrt{Hz}}$ was attained using a analyzing area of $28~\mathrm{\mu m^2}$ with $t_{0}=300~\mathrm{ms}$, $\Delta= -1.257~\mathrm{GHz}$, and $\Phi_{p} = 8.3\times10^{3}~\mathrm{\mu m^{-2}}$. 
This field sensitivity was comparable to those of previous BEC magnetometers \cite{Wildermuth2005, Vengalattore2007, Jasperse2017, Muessel2014, Eto2013}.
More importantly, the sensitivity was 
obtained with only $3 \times 10^4$ atoms in the small measurement area.  
We improved the sensitivity per measurement area by about a factor of 4 compared with the previous record for BEC magnetometers \cite{Vengalattore2007}. 
The magnetic field energy resolution was evaluated to be $58\hbar$ and is comparable to reported values for SQUIDs \cite{Awschalom1988, Wakai1988} at low frequency and for the most sensitive atomic magnetometer \cite{Dang2010}.
The improvement was achieved by optimizing the probe parameters, which reduced the light-induced signal decay and suppressed the photon shot noise, a dominant sensitivity limitation in BEC magnetometers \cite{Vengalattore2007,Jasperse2017}.

With practically feasible improvements, a sensitivity below $\mathrm{pT/\sqrt{Hz}}$ is obtainable.
As we operated the magnetometer with a low duty cycle of $5 \times 10^{-3}$, the sensitivity can be greatly improved by the fast production of a BEC \cite{Boyd2006} or extending the magnetic coherence time using a single-mode BEC \cite{Jasperse2017,Palacios2018}. For AC field detection, bang--bang control is also useful \cite{Eto2014}. 

The fundamental ultimate limitation of magnetometry should be evaluated based on the \textit{essential} sensitivity with a full duty cycle and negligible magnetic decoherence, which ideally holds for a spinor BEC.
The essential energy resolution for our BEC magnetometry is estimated to be 0.3 $\hbar$, which is better than the energy resolutions for state-of-the-art solid-state magnetometers including diamond sensors using spin ensembles \cite{Wolf2015}, and even surpasses those for micro/nano SQUIDs \cite{Wakai1988,Awschalom1988,Muck2001}, the limitation of which has been shown to be $\approx \hbar$ \cite{Koch1981, Koch1980}.
Our result shows that BECs may be potentially superior to SQUIDs for spatial magnetometers.

In conclusion, we realized sensitive spin detection of a BEC with TPPCI.
We optimized the probe parameters, making use of the ability in TPPCI to track atom and spin density, and achieved a Larmor-phase sensitivity of 5.6~mrad over a measurement area of $142~\mathrm{\mu m^{2}}$.
We obtained a dc-magnetic field sensitivity of $7.7~\mathrm{pT/\sqrt{Hz}}$ over a region of $28~\mathrm{\mu m^{2}}$.
Furthermore, the TPPCI used here is compatible with spin squeezing, and planar squeezing in particular \cite{Puentes2013,Colangelo2017}.
Although BECs are potentially promising for achieving a high squeezing level due to their large optical density, the signal decay due to the probe has inhibited measurement-based spin squeezing in a BEC.
Our result demonstrates that a large atom--light interaction can be obtained by optimizing the probe parameters.
Furthermore, the performance of our magnetometry method, which is comparable to that of state-of-the-art magnetometers, shows that true quantum-enhanced magnetometry with a sensitivity better than that possible with classical methods is feasible.

\begin{acknowledgements}
This work was supported by MEXT Quantum Leap Flagship Program (MEXT Q-LEAP) Grant Number JPMXS0118070326.
\end{acknowledgements}


\begin{thebibliography}{41}%
\makeatletter
\providecommand \@ifxundefined [1]{%
 \@ifx{#1\undefined}
}%
\providecommand \@ifnum [1]{%
 \ifnum #1\expandafter \@firstoftwo
 \else \expandafter \@secondoftwo
 \fi
}%
\providecommand \@ifx [1]{%
 \ifx #1\expandafter \@firstoftwo
 \else \expandafter \@secondoftwo
 \fi
}%
\providecommand \natexlab [1]{#1}%
\providecommand \enquote  [1]{``#1''}%
\providecommand \bibnamefont  [1]{#1}%
\providecommand \bibfnamefont [1]{#1}%
\providecommand \citenamefont [1]{#1}%
\providecommand \href@noop [0]{\@secondoftwo}%
\providecommand \href [0]{\begingroup \@sanitize@url \@href}%
\providecommand \@href[1]{\@@startlink{#1}\@@href}%
\providecommand \@@href[1]{\endgroup#1\@@endlink}%
\providecommand \@sanitize@url [0]{\catcode `\\12\catcode `\$12\catcode
  `\&12\catcode `\#12\catcode `\^12\catcode `\_12\catcode `\%12\relax}%
\providecommand \@@startlink[1]{}%
\providecommand \@@endlink[0]{}%
\providecommand \url  [0]{\begingroup\@sanitize@url \@url }%
\providecommand \@url [1]{\endgroup\@href {#1}{\urlprefix }}%
\providecommand \urlprefix  [0]{URL }%
\providecommand \Eprint [0]{\href }%
\providecommand \doibase [0]{https://doi.org/}%
\providecommand \selectlanguage [0]{\@gobble}%
\providecommand \bibinfo  [0]{\@secondoftwo}%
\providecommand \bibfield  [0]{\@secondoftwo}%
\providecommand \translation [1]{[#1]}%
\providecommand \BibitemOpen [0]{}%
\providecommand \bibitemStop [0]{}%
\providecommand \bibitemNoStop [0]{.\EOS\space}%
\providecommand \EOS [0]{\spacefactor3000\relax}%
\providecommand \BibitemShut  [1]{\csname bibitem#1\endcsname}%
\let\auto@bib@innerbib\@empty
%</preamble>
\bibitem [{\citenamefont {Kominis}\ \emph {et~al.}(2003)\citenamefont
  {Kominis}, \citenamefont {Kornack}, \citenamefont {Allred},\ and\
  \citenamefont {Romalis}}]{Kominis2003}%
  \BibitemOpen
  \bibfield  {author} {\bibinfo {author} {\bibfnamefont {I.~K.}\ \bibnamefont
  {Kominis}}, \bibinfo {author} {\bibfnamefont {T.~W.}\ \bibnamefont
  {Kornack}}, \bibinfo {author} {\bibfnamefont {J.~C.}\ \bibnamefont
  {Allred}},\ and\ \bibinfo {author} {\bibfnamefont {M.~V.}\ \bibnamefont
  {Romalis}},\ }\href {https://doi.org/10.1038/nature01484} {\bibfield
  {journal} {\bibinfo  {journal} {Nature}\ }\textbf {\bibinfo {volume} {422}},\
  \bibinfo {pages} {596} (\bibinfo {year} {2003})}\BibitemShut {NoStop}%
\bibitem [{\citenamefont {Dang}\ \emph {et~al.}(2010)\citenamefont {Dang},
  \citenamefont {Maloof},\ and\ \citenamefont {Romalis}}]{Dang2010}%
  \BibitemOpen
  \bibfield  {author} {\bibinfo {author} {\bibfnamefont {H.~B.}\ \bibnamefont
  {Dang}}, \bibinfo {author} {\bibfnamefont {A.~C.}\ \bibnamefont {Maloof}},\
  and\ \bibinfo {author} {\bibfnamefont {M.~V.}\ \bibnamefont {Romalis}},\
  }\href {https://doi.org/10.1063/1.3491215} {\bibfield  {journal} {\bibinfo
  {journal} {Applied Physics Letters}\ }\textbf {\bibinfo {volume} {97}},\
  \bibinfo {pages} {151110} (\bibinfo {year} {2010})}\BibitemShut {NoStop}%
\bibitem [{\citenamefont {Schmelz}\ \emph {et~al.}(2011)\citenamefont
  {Schmelz}, \citenamefont {Stolz}, \citenamefont {Zakosarenko}, \citenamefont
  {Sch{\"{o}}nau}, \citenamefont {Anders}, \citenamefont {Fritzsch},
  \citenamefont {M{\"{u}}ck},\ and\ \citenamefont {Meyer}}]{Schmelz2011}%
  \BibitemOpen
  \bibfield  {author} {\bibinfo {author} {\bibfnamefont {M.}~\bibnamefont
  {Schmelz}}, \bibinfo {author} {\bibfnamefont {R.}~\bibnamefont {Stolz}},
  \bibinfo {author} {\bibfnamefont {V.}~\bibnamefont {Zakosarenko}}, \bibinfo
  {author} {\bibfnamefont {T.}~\bibnamefont {Sch{\"{o}}nau}}, \bibinfo {author}
  {\bibfnamefont {S.}~\bibnamefont {Anders}}, \bibinfo {author} {\bibfnamefont
  {L.}~\bibnamefont {Fritzsch}}, \bibinfo {author} {\bibfnamefont
  {M.}~\bibnamefont {M{\"{u}}ck}},\ and\ \bibinfo {author} {\bibfnamefont
  {H.-G.}\ \bibnamefont {Meyer}},\ }\href
  {https://doi.org/10.1088/0953-2048/24/6/065009} {\bibfield  {journal}
  {\bibinfo  {journal} {Superconductor Science and Technology}\ }\textbf
  {\bibinfo {volume} {24}},\ \bibinfo {pages} {065009} (\bibinfo {year}
  {2011})}\BibitemShut {NoStop}%
\bibitem [{\citenamefont {Baumgart}\ \emph {et~al.}(2016)\citenamefont
  {Baumgart}, \citenamefont {Cai}, \citenamefont {Retzker}, \citenamefont
  {Plenio},\ and\ \citenamefont {Wunderlich}}]{Baumgart2016}%
  \BibitemOpen
  \bibfield  {author} {\bibinfo {author} {\bibfnamefont {I.}~\bibnamefont
  {Baumgart}}, \bibinfo {author} {\bibfnamefont {J.-M.}\ \bibnamefont {Cai}},
  \bibinfo {author} {\bibfnamefont {A.}~\bibnamefont {Retzker}}, \bibinfo
  {author} {\bibfnamefont {M.~B.}\ \bibnamefont {Plenio}},\ and\ \bibinfo
  {author} {\bibfnamefont {C.}~\bibnamefont {Wunderlich}},\ }\href
  {https://doi.org/10.1103/PhysRevLett.116.240801} {\bibfield  {journal}
  {\bibinfo  {journal} {Physical Review Letters}\ }\textbf {\bibinfo {volume}
  {116}},\ \bibinfo {pages} {240801} (\bibinfo {year} {2016})}\BibitemShut
  {NoStop}%
\bibitem [{\citenamefont {Balasubramanian}\ \emph {et~al.}(2009)\citenamefont
  {Balasubramanian}, \citenamefont {Neumann}, \citenamefont {Twitchen},
  \citenamefont {Markham}, \citenamefont {Kolesov}, \citenamefont {Mizuochi},
  \citenamefont {Isoya}, \citenamefont {Achard}, \citenamefont {Beck},
  \citenamefont {Tissler}, \citenamefont {Jacques}, \citenamefont {Hemmer},
  \citenamefont {Jelezko},\ and\ \citenamefont
  {Wrachtrup}}]{Balasubramanian2009}%
  \BibitemOpen
  \bibfield  {author} {\bibinfo {author} {\bibfnamefont {G.}~\bibnamefont
  {Balasubramanian}}, \bibinfo {author} {\bibfnamefont {P.}~\bibnamefont
  {Neumann}}, \bibinfo {author} {\bibfnamefont {D.}~\bibnamefont {Twitchen}},
  \bibinfo {author} {\bibfnamefont {M.}~\bibnamefont {Markham}}, \bibinfo
  {author} {\bibfnamefont {R.}~\bibnamefont {Kolesov}}, \bibinfo {author}
  {\bibfnamefont {N.}~\bibnamefont {Mizuochi}}, \bibinfo {author}
  {\bibfnamefont {J.}~\bibnamefont {Isoya}}, \bibinfo {author} {\bibfnamefont
  {J.}~\bibnamefont {Achard}}, \bibinfo {author} {\bibfnamefont
  {J.}~\bibnamefont {Beck}}, \bibinfo {author} {\bibfnamefont {J.}~\bibnamefont
  {Tissler}}, \bibinfo {author} {\bibfnamefont {V.}~\bibnamefont {Jacques}},
  \bibinfo {author} {\bibfnamefont {P.~R.}\ \bibnamefont {Hemmer}}, \bibinfo
  {author} {\bibfnamefont {F.}~\bibnamefont {Jelezko}},\ and\ \bibinfo {author}
  {\bibfnamefont {J.}~\bibnamefont {Wrachtrup}},\ }\href
  {https://doi.org/10.1038/nmat2420} {\bibfield  {journal} {\bibinfo  {journal}
  {Nature Materials}\ }\textbf {\bibinfo {volume} {8}},\ \bibinfo {pages} {383}
  (\bibinfo {year} {2009})}\BibitemShut {NoStop}%
\bibitem [{\citenamefont {El{\'{i}}asson}\ \emph {et~al.}(2019)\citenamefont
  {El{\'{i}}asson}, \citenamefont {Heck}, \citenamefont {Laustsen},
  \citenamefont {Napolitano}, \citenamefont {M{\"{u}}ller}, \citenamefont
  {Bason}, \citenamefont {Arlt},\ and\ \citenamefont {Sherson}}]{Eliasson2019}%
  \BibitemOpen
  \bibfield  {author} {\bibinfo {author} {\bibfnamefont {O.}~\bibnamefont
  {El{\'{i}}asson}}, \bibinfo {author} {\bibfnamefont {R.}~\bibnamefont
  {Heck}}, \bibinfo {author} {\bibfnamefont {J.~S.}\ \bibnamefont {Laustsen}},
  \bibinfo {author} {\bibfnamefont {M.}~\bibnamefont {Napolitano}}, \bibinfo
  {author} {\bibfnamefont {R.}~\bibnamefont {M{\"{u}}ller}}, \bibinfo {author}
  {\bibfnamefont {M.~G.}\ \bibnamefont {Bason}}, \bibinfo {author}
  {\bibfnamefont {J.~J.}\ \bibnamefont {Arlt}},\ and\ \bibinfo {author}
  {\bibfnamefont {J.~F.}\ \bibnamefont {Sherson}},\ }\href
  {https://doi.org/10.1088/1361-6455/ab0bd6} {\bibfield  {journal} {\bibinfo
  {journal} {Journal of Physics B: Atomic, Molecular and Optical Physics}\
  }\textbf {\bibinfo {volume} {52}},\ \bibinfo {pages} {075003} (\bibinfo
  {year} {2019})}\BibitemShut {NoStop}%
\bibitem [{\citenamefont {Wildermuth}\ \emph {et~al.}(2005)\citenamefont
  {Wildermuth}, \citenamefont {Hofferberth}, \citenamefont {Lesanovsky},
  \citenamefont {Haller}, \citenamefont {Andersson}, \citenamefont {Groth},
  \citenamefont {Bar-Joseph}, \citenamefont {Kr{\"{u}}ger},\ and\ \citenamefont
  {Schmiedmayer}}]{Wildermuth2005}%
  \BibitemOpen
  \bibfield  {author} {\bibinfo {author} {\bibfnamefont {S.}~\bibnamefont
  {Wildermuth}}, \bibinfo {author} {\bibfnamefont {S.}~\bibnamefont
  {Hofferberth}}, \bibinfo {author} {\bibfnamefont {I.}~\bibnamefont
  {Lesanovsky}}, \bibinfo {author} {\bibfnamefont {E.}~\bibnamefont {Haller}},
  \bibinfo {author} {\bibfnamefont {L.~M.}\ \bibnamefont {Andersson}}, \bibinfo
  {author} {\bibfnamefont {S.}~\bibnamefont {Groth}}, \bibinfo {author}
  {\bibfnamefont {I.}~\bibnamefont {Bar-Joseph}}, \bibinfo {author}
  {\bibfnamefont {P.}~\bibnamefont {Kr{\"{u}}ger}},\ and\ \bibinfo {author}
  {\bibfnamefont {J.}~\bibnamefont {Schmiedmayer}},\ }\href
  {https://doi.org/10.1038/435440a} {\bibfield  {journal} {\bibinfo  {journal}
  {Nature}\ }\textbf {\bibinfo {volume} {435}},\ \bibinfo {pages} {440}
  (\bibinfo {year} {2005})}\BibitemShut {NoStop}%
\bibitem [{\citenamefont {Vengalattore}\ \emph {et~al.}(2007)\citenamefont
  {Vengalattore}, \citenamefont {Higbie}, \citenamefont {Leslie}, \citenamefont
  {Guzman}, \citenamefont {Sadler},\ and\ \citenamefont
  {Stamper-Kurn}}]{Vengalattore2007}%
  \BibitemOpen
  \bibfield  {author} {\bibinfo {author} {\bibfnamefont {M.}~\bibnamefont
  {Vengalattore}}, \bibinfo {author} {\bibfnamefont {J.~M.}\ \bibnamefont
  {Higbie}}, \bibinfo {author} {\bibfnamefont {S.~R.}\ \bibnamefont {Leslie}},
  \bibinfo {author} {\bibfnamefont {J.}~\bibnamefont {Guzman}}, \bibinfo
  {author} {\bibfnamefont {L.~E.}\ \bibnamefont {Sadler}},\ and\ \bibinfo
  {author} {\bibfnamefont {D.~M.}\ \bibnamefont {Stamper-Kurn}},\ }\href
  {https://doi.org/10.1103/PhysRevLett.98.200801} {\bibfield  {journal}
  {\bibinfo  {journal} {Physical Review Letters}\ }\textbf {\bibinfo {volume}
  {98}},\ \bibinfo {pages} {200801} (\bibinfo {year} {2007})}\BibitemShut
  {NoStop}%
\bibitem [{\citenamefont {Jasperse}\ \emph {et~al.}(2017)\citenamefont
  {Jasperse}, \citenamefont {Kewming}, \citenamefont {Fischer}, \citenamefont
  {Pakkiam}, \citenamefont {Anderson},\ and\ \citenamefont
  {Turner}}]{Jasperse2017}%
  \BibitemOpen
  \bibfield  {author} {\bibinfo {author} {\bibfnamefont {M.}~\bibnamefont
  {Jasperse}}, \bibinfo {author} {\bibfnamefont {M.~J.}\ \bibnamefont
  {Kewming}}, \bibinfo {author} {\bibfnamefont {S.~N.}\ \bibnamefont
  {Fischer}}, \bibinfo {author} {\bibfnamefont {P.}~\bibnamefont {Pakkiam}},
  \bibinfo {author} {\bibfnamefont {R.~P.}\ \bibnamefont {Anderson}},\ and\
  \bibinfo {author} {\bibfnamefont {L.~D.}\ \bibnamefont {Turner}},\ }\href
  {https://doi.org/10.1103/PhysRevA.96.063402} {\bibfield  {journal} {\bibinfo
  {journal} {Physical Review A}\ }\textbf {\bibinfo {volume} {96}},\ \bibinfo
  {pages} {063402} (\bibinfo {year} {2017})}\BibitemShut {NoStop}%
\bibitem [{\citenamefont {Palacios}\ \emph {et~al.}(2018)\citenamefont
  {Palacios}, \citenamefont {Coop}, \citenamefont {Gomez}, \citenamefont
  {Vanderbruggen}, \citenamefont {{Martinez de Escobar}}, \citenamefont
  {Jasperse},\ and\ \citenamefont {Mitchell}}]{Palacios2018}%
  \BibitemOpen
  \bibfield  {author} {\bibinfo {author} {\bibfnamefont {S.}~\bibnamefont
  {Palacios}}, \bibinfo {author} {\bibfnamefont {S.}~\bibnamefont {Coop}},
  \bibinfo {author} {\bibfnamefont {P.}~\bibnamefont {Gomez}}, \bibinfo
  {author} {\bibfnamefont {T.}~\bibnamefont {Vanderbruggen}}, \bibinfo {author}
  {\bibfnamefont {Y.~N.}\ \bibnamefont {{Martinez de Escobar}}}, \bibinfo
  {author} {\bibfnamefont {M.}~\bibnamefont {Jasperse}},\ and\ \bibinfo
  {author} {\bibfnamefont {M.~W.}\ \bibnamefont {Mitchell}},\ }\href
  {https://doi.org/10.1088/1367-2630/aab2a0} {\bibfield  {journal} {\bibinfo
  {journal} {New Journal of Physics}\ }\textbf {\bibinfo {volume} {20}},\
  \bibinfo {pages} {053008} (\bibinfo {year} {2018})}\BibitemShut {NoStop}%
\bibitem [{\citenamefont {H{\"{a}}m{\"{a}}l{\"{a}}inen}\ \emph
  {et~al.}(1993)\citenamefont {H{\"{a}}m{\"{a}}l{\"{a}}inen}, \citenamefont
  {Hari}, \citenamefont {Ilmoniemi}, \citenamefont {Knuutila},\ and\
  \citenamefont {Lounasmaa}}]{Hamalainen1993}%
  \BibitemOpen
  \bibfield  {author} {\bibinfo {author} {\bibfnamefont {M.}~\bibnamefont
  {H{\"{a}}m{\"{a}}l{\"{a}}inen}}, \bibinfo {author} {\bibfnamefont
  {R.}~\bibnamefont {Hari}}, \bibinfo {author} {\bibfnamefont {R.~J.}\
  \bibnamefont {Ilmoniemi}}, \bibinfo {author} {\bibfnamefont {J.}~\bibnamefont
  {Knuutila}},\ and\ \bibinfo {author} {\bibfnamefont {O.~V.}\ \bibnamefont
  {Lounasmaa}},\ }\href {https://doi.org/10.1103/RevModPhys.65.413} {\bibfield
  {journal} {\bibinfo  {journal} {Reviews of Modern Physics}\ }\textbf
  {\bibinfo {volume} {65}},\ \bibinfo {pages} {413} (\bibinfo {year}
  {1993})}\BibitemShut {NoStop}%
\bibitem [{\citenamefont {Yang}\ \emph {et~al.}(2020)\citenamefont {Yang},
  \citenamefont {Taylor}, \citenamefont {Edkins}, \citenamefont {Palmstrom},
  \citenamefont {Fisher},\ and\ \citenamefont {Lev}}]{Yang2020}%
  \BibitemOpen
  \bibfield  {author} {\bibinfo {author} {\bibfnamefont {F.}~\bibnamefont
  {Yang}}, \bibinfo {author} {\bibfnamefont {S.~F.}\ \bibnamefont {Taylor}},
  \bibinfo {author} {\bibfnamefont {S.~D.}\ \bibnamefont {Edkins}}, \bibinfo
  {author} {\bibfnamefont {J.~C.}\ \bibnamefont {Palmstrom}}, \bibinfo {author}
  {\bibfnamefont {I.~R.}\ \bibnamefont {Fisher}},\ and\ \bibinfo {author}
  {\bibfnamefont {B.~L.}\ \bibnamefont {Lev}},\ }\href
  {https://doi.org/10.1038/s41567-020-0826-8} {\bibfield  {journal} {\bibinfo
  {journal} {Nature Physics}\ }\textbf {\bibinfo {volume} {16}},\ \bibinfo
  {pages} {514} (\bibinfo {year} {2020})}\BibitemShut {NoStop}%
\bibitem [{\citenamefont {Hu}\ \emph {et~al.}(2020)\citenamefont {Hu},
  \citenamefont {Iwata}, \citenamefont {Mohammadi}, \citenamefont {Silletta},
  \citenamefont {Wickenbrock}, \citenamefont {Blanchard}, \citenamefont
  {Budker},\ and\ \citenamefont {Jerschow}}]{Hu2020}%
  \BibitemOpen
  \bibfield  {author} {\bibinfo {author} {\bibfnamefont {Y.}~\bibnamefont
  {Hu}}, \bibinfo {author} {\bibfnamefont {G.~Z.}\ \bibnamefont {Iwata}},
  \bibinfo {author} {\bibfnamefont {M.}~\bibnamefont {Mohammadi}}, \bibinfo
  {author} {\bibfnamefont {E.~V.}\ \bibnamefont {Silletta}}, \bibinfo {author}
  {\bibfnamefont {A.}~\bibnamefont {Wickenbrock}}, \bibinfo {author}
  {\bibfnamefont {J.~W.}\ \bibnamefont {Blanchard}}, \bibinfo {author}
  {\bibfnamefont {D.}~\bibnamefont {Budker}},\ and\ \bibinfo {author}
  {\bibfnamefont {A.}~\bibnamefont {Jerschow}},\ }\href
  {https://doi.org/10.1073/pnas.1917172117} {\bibfield  {journal} {\bibinfo
  {journal} {Proceedings of the National Academy of Sciences}\ }\textbf
  {\bibinfo {volume} {117}},\ \bibinfo {pages} {10667} (\bibinfo {year}
  {2020})}\BibitemShut {NoStop}%
\bibitem [{\citenamefont {Mitchell}\ and\ \citenamefont {{Palacios
  Alvarez}}(2020)}]{Mitchell2020}%
  \BibitemOpen
  \bibfield  {author} {\bibinfo {author} {\bibfnamefont {M.~W.}\ \bibnamefont
  {Mitchell}}\ and\ \bibinfo {author} {\bibfnamefont {S.}~\bibnamefont
  {{Palacios Alvarez}}},\ }\href {https://doi.org/10.1103/RevModPhys.92.021001}
  {\bibfield  {journal} {\bibinfo  {journal} {Reviews of Modern Physics}\
  }\textbf {\bibinfo {volume} {92}},\ \bibinfo {pages} {021001} (\bibinfo
  {year} {2020})}\BibitemShut {NoStop}%
\bibitem [{\citenamefont {Sewell}\ \emph {et~al.}(2012)\citenamefont {Sewell},
  \citenamefont {Koschorreck}, \citenamefont {Napolitano}, \citenamefont
  {Dubost}, \citenamefont {Behbood},\ and\ \citenamefont
  {Mitchell}}]{Sewell2012}%
  \BibitemOpen
  \bibfield  {author} {\bibinfo {author} {\bibfnamefont {R.~J.}\ \bibnamefont
  {Sewell}}, \bibinfo {author} {\bibfnamefont {M.}~\bibnamefont {Koschorreck}},
  \bibinfo {author} {\bibfnamefont {M.}~\bibnamefont {Napolitano}}, \bibinfo
  {author} {\bibfnamefont {B.}~\bibnamefont {Dubost}}, \bibinfo {author}
  {\bibfnamefont {N.}~\bibnamefont {Behbood}},\ and\ \bibinfo {author}
  {\bibfnamefont {M.~W.}\ \bibnamefont {Mitchell}},\ }\href
  {https://doi.org/10.1103/PhysRevLett.109.253605} {\bibfield  {journal}
  {\bibinfo  {journal} {Physical Review Letters}\ }\textbf {\bibinfo {volume}
  {109}},\ \bibinfo {pages} {253605} (\bibinfo {year} {2012})}\BibitemShut
  {NoStop}%
\bibitem [{\citenamefont {Muessel}\ \emph {et~al.}(2014)\citenamefont
  {Muessel}, \citenamefont {Strobel}, \citenamefont {Linnemann}, \citenamefont
  {Hume},\ and\ \citenamefont {Oberthaler}}]{Muessel2014}%
  \BibitemOpen
  \bibfield  {author} {\bibinfo {author} {\bibfnamefont {W.}~\bibnamefont
  {Muessel}}, \bibinfo {author} {\bibfnamefont {H.}~\bibnamefont {Strobel}},
  \bibinfo {author} {\bibfnamefont {D.}~\bibnamefont {Linnemann}}, \bibinfo
  {author} {\bibfnamefont {D.~B.}\ \bibnamefont {Hume}},\ and\ \bibinfo
  {author} {\bibfnamefont {M.~K.}\ \bibnamefont {Oberthaler}},\ }\href
  {https://doi.org/10.1103/PhysRevLett.113.103004} {\bibfield  {journal}
  {\bibinfo  {journal} {Physical Review Letters}\ }\textbf {\bibinfo {volume}
  {113}},\ \bibinfo {pages} {103004} (\bibinfo {year} {2014})}\BibitemShut
  {NoStop}%
\bibitem [{\citenamefont {{Martin Ciurana}}\ \emph {et~al.}(2017)\citenamefont
  {{Martin Ciurana}}, \citenamefont {Colangelo}, \citenamefont
  {Slodi{\v{c}}ka}, \citenamefont {Sewell},\ and\ \citenamefont
  {Mitchell}}]{MartinCiurana2017}%
  \BibitemOpen
  \bibfield  {author} {\bibinfo {author} {\bibfnamefont {F.}~\bibnamefont
  {{Martin Ciurana}}}, \bibinfo {author} {\bibfnamefont {G.}~\bibnamefont
  {Colangelo}}, \bibinfo {author} {\bibfnamefont {L.}~\bibnamefont
  {Slodi{\v{c}}ka}}, \bibinfo {author} {\bibfnamefont {R.~J.}\ \bibnamefont
  {Sewell}},\ and\ \bibinfo {author} {\bibfnamefont {M.~W.}\ \bibnamefont
  {Mitchell}},\ }\href {https://doi.org/10.1103/PhysRevLett.119.043603}
  {\bibfield  {journal} {\bibinfo  {journal} {Physical Review Letters}\
  }\textbf {\bibinfo {volume} {119}},\ \bibinfo {pages} {043603} (\bibinfo
  {year} {2017})}\BibitemShut {NoStop}%
\bibitem [{\citenamefont {Colangelo}\ \emph {et~al.}(2017)\citenamefont
  {Colangelo}, \citenamefont {Ciurana}, \citenamefont {Bianchet}, \citenamefont
  {Sewell},\ and\ \citenamefont {Mitchell}}]{Colangelo2017}%
  \BibitemOpen
  \bibfield  {author} {\bibinfo {author} {\bibfnamefont {G.}~\bibnamefont
  {Colangelo}}, \bibinfo {author} {\bibfnamefont {F.~M.}\ \bibnamefont
  {Ciurana}}, \bibinfo {author} {\bibfnamefont {L.~C.}\ \bibnamefont
  {Bianchet}}, \bibinfo {author} {\bibfnamefont {R.~J.}\ \bibnamefont
  {Sewell}},\ and\ \bibinfo {author} {\bibfnamefont {M.~W.}\ \bibnamefont
  {Mitchell}},\ }\href {https://doi.org/10.1038/nature21434} {\bibfield
  {journal} {\bibinfo  {journal} {Nature}\ }\textbf {\bibinfo {volume} {543}},\
  \bibinfo {pages} {525} (\bibinfo {year} {2017})}\BibitemShut {NoStop}%
\bibitem [{\citenamefont {Chalopin}\ \emph {et~al.}(2018)\citenamefont
  {Chalopin}, \citenamefont {Bouazza}, \citenamefont {Evrard}, \citenamefont
  {Makhalov}, \citenamefont {Dreon}, \citenamefont {Dalibard}, \citenamefont
  {Sidorenkov},\ and\ \citenamefont {Nascimbene}}]{Chalopin2018}%
  \BibitemOpen
  \bibfield  {author} {\bibinfo {author} {\bibfnamefont {T.}~\bibnamefont
  {Chalopin}}, \bibinfo {author} {\bibfnamefont {C.}~\bibnamefont {Bouazza}},
  \bibinfo {author} {\bibfnamefont {A.}~\bibnamefont {Evrard}}, \bibinfo
  {author} {\bibfnamefont {V.}~\bibnamefont {Makhalov}}, \bibinfo {author}
  {\bibfnamefont {D.}~\bibnamefont {Dreon}}, \bibinfo {author} {\bibfnamefont
  {J.}~\bibnamefont {Dalibard}}, \bibinfo {author} {\bibfnamefont {L.~A.}\
  \bibnamefont {Sidorenkov}},\ and\ \bibinfo {author} {\bibfnamefont
  {S.}~\bibnamefont {Nascimbene}},\ }\href
  {https://doi.org/10.1038/s41467-018-07433-1} {\bibfield  {journal} {\bibinfo
  {journal} {Nature Communications}\ }\textbf {\bibinfo {volume} {9}},\
  \bibinfo {pages} {4955} (\bibinfo {year} {2018})}\BibitemShut {NoStop}%
\bibitem [{\citenamefont {Evrard}\ \emph {et~al.}(2019)\citenamefont {Evrard},
  \citenamefont {Makhalov}, \citenamefont {Chalopin}, \citenamefont
  {Sidorenkov}, \citenamefont {Dalibard}, \citenamefont {Lopes},\ and\
  \citenamefont {Nascimbene}}]{Evrard2019}%
  \BibitemOpen
  \bibfield  {author} {\bibinfo {author} {\bibfnamefont {A.}~\bibnamefont
  {Evrard}}, \bibinfo {author} {\bibfnamefont {V.}~\bibnamefont {Makhalov}},
  \bibinfo {author} {\bibfnamefont {T.}~\bibnamefont {Chalopin}}, \bibinfo
  {author} {\bibfnamefont {L.~A.}\ \bibnamefont {Sidorenkov}}, \bibinfo
  {author} {\bibfnamefont {J.}~\bibnamefont {Dalibard}}, \bibinfo {author}
  {\bibfnamefont {R.}~\bibnamefont {Lopes}},\ and\ \bibinfo {author}
  {\bibfnamefont {S.}~\bibnamefont {Nascimbene}},\ }\href
  {https://doi.org/10.1103/PhysRevLett.122.173601} {\bibfield  {journal}
  {\bibinfo  {journal} {Physical Review Letters}\ }\textbf {\bibinfo {volume}
  {122}},\ \bibinfo {pages} {173601} (\bibinfo {year} {2019})}\BibitemShut
  {NoStop}%
\bibitem [{\citenamefont {Smith}\ \emph {et~al.}(2004)\citenamefont {Smith},
  \citenamefont {Chaudhury}, \citenamefont {Silberfarb}, \citenamefont
  {Deutsch},\ and\ \citenamefont {Jessen}}]{Smith2004}%
  \BibitemOpen
  \bibfield  {author} {\bibinfo {author} {\bibfnamefont {G.~A.}\ \bibnamefont
  {Smith}}, \bibinfo {author} {\bibfnamefont {S.}~\bibnamefont {Chaudhury}},
  \bibinfo {author} {\bibfnamefont {A.}~\bibnamefont {Silberfarb}}, \bibinfo
  {author} {\bibfnamefont {I.~H.}\ \bibnamefont {Deutsch}},\ and\ \bibinfo
  {author} {\bibfnamefont {P.~S.}\ \bibnamefont {Jessen}},\ }\href
  {https://doi.org/10.1103/PhysRevLett.93.163602} {\bibfield  {journal}
  {\bibinfo  {journal} {Physical Review Letters}\ }\textbf {\bibinfo {volume}
  {93}},\ \bibinfo {pages} {163602} (\bibinfo {year} {2004})}\BibitemShut
  {NoStop}%
\bibitem [{\citenamefont {Deutsch}\ and\ \citenamefont
  {Jessen}(2010)}]{Deutsch2010}%
  \BibitemOpen
  \bibfield  {author} {\bibinfo {author} {\bibfnamefont {I.~H.}\ \bibnamefont
  {Deutsch}}\ and\ \bibinfo {author} {\bibfnamefont {P.~S.}\ \bibnamefont
  {Jessen}},\ }\href {https://doi.org/10.1016/j.optcom.2009.10.059} {\bibfield
  {journal} {\bibinfo  {journal} {Optics Communications}\ }\textbf {\bibinfo
  {volume} {283}},\ \bibinfo {pages} {681} (\bibinfo {year}
  {2010})}\BibitemShut {NoStop}%
\bibitem [{\citenamefont {Colangelo}\ \emph {et~al.}(2013)\citenamefont
  {Colangelo}, \citenamefont {Sewell}, \citenamefont {Behbood}, \citenamefont
  {Ciurana}, \citenamefont {Triginer},\ and\ \citenamefont
  {Mitchell}}]{Colangelo2013}%
  \BibitemOpen
  \bibfield  {author} {\bibinfo {author} {\bibfnamefont {G.}~\bibnamefont
  {Colangelo}}, \bibinfo {author} {\bibfnamefont {R.~J.}\ \bibnamefont
  {Sewell}}, \bibinfo {author} {\bibfnamefont {N.}~\bibnamefont {Behbood}},
  \bibinfo {author} {\bibfnamefont {F.~M.}\ \bibnamefont {Ciurana}}, \bibinfo
  {author} {\bibfnamefont {G.}~\bibnamefont {Triginer}},\ and\ \bibinfo
  {author} {\bibfnamefont {M.~W.}\ \bibnamefont {Mitchell}},\ }\href
  {https://doi.org/10.1088/1367-2630/15/10/103007} {\bibfield  {journal}
  {\bibinfo  {journal} {New Journal of Physics}\ }\textbf {\bibinfo {volume}
  {15}},\ \bibinfo {pages} {103007} (\bibinfo {year} {2013})}\BibitemShut
  {NoStop}%
\bibitem [{\citenamefont {Urvoy}\ \emph {et~al.}(2019)\citenamefont {Urvoy},
  \citenamefont {Vendeiro}, \citenamefont {Ramette}, \citenamefont
  {Adiyatullin},\ and\ \citenamefont {Vuleti{\'{c}}}}]{Urvoy2019}%
  \BibitemOpen
  \bibfield  {author} {\bibinfo {author} {\bibfnamefont {A.}~\bibnamefont
  {Urvoy}}, \bibinfo {author} {\bibfnamefont {Z.}~\bibnamefont {Vendeiro}},
  \bibinfo {author} {\bibfnamefont {J.}~\bibnamefont {Ramette}}, \bibinfo
  {author} {\bibfnamefont {A.}~\bibnamefont {Adiyatullin}},\ and\ \bibinfo
  {author} {\bibfnamefont {V.}~\bibnamefont {Vuleti{\'{c}}}},\ }\href
  {https://doi.org/10.1103/PhysRevLett.122.203202} {\bibfield  {journal}
  {\bibinfo  {journal} {Physical Review Letters}\ }\textbf {\bibinfo {volume}
  {122}},\ \bibinfo {pages} {203202} (\bibinfo {year} {2019})}\BibitemShut
  {NoStop}%
\bibitem [{\citenamefont {Higbie}\ \emph {et~al.}(2005)\citenamefont {Higbie},
  \citenamefont {Sadler}, \citenamefont {Inouye}, \citenamefont {Chikkatur},
  \citenamefont {Leslie}, \citenamefont {Moore}, \citenamefont {Savalli},\ and\
  \citenamefont {Stamper-Kurn}}]{Higbie2005}%
  \BibitemOpen
  \bibfield  {author} {\bibinfo {author} {\bibfnamefont {J.~M.}\ \bibnamefont
  {Higbie}}, \bibinfo {author} {\bibfnamefont {L.~E.}\ \bibnamefont {Sadler}},
  \bibinfo {author} {\bibfnamefont {S.}~\bibnamefont {Inouye}}, \bibinfo
  {author} {\bibfnamefont {A.~P.}\ \bibnamefont {Chikkatur}}, \bibinfo {author}
  {\bibfnamefont {S.~R.}\ \bibnamefont {Leslie}}, \bibinfo {author}
  {\bibfnamefont {K.~L.}\ \bibnamefont {Moore}}, \bibinfo {author}
  {\bibfnamefont {V.}~\bibnamefont {Savalli}},\ and\ \bibinfo {author}
  {\bibfnamefont {D.~M.}\ \bibnamefont {Stamper-Kurn}},\ }\href
  {https://doi.org/10.1103/PhysRevLett.95.050401} {\bibfield  {journal}
  {\bibinfo  {journal} {Physical Review Letters}\ }\textbf {\bibinfo {volume}
  {95}},\ \bibinfo {pages} {050401} (\bibinfo {year} {2005})}\BibitemShut
  {NoStop}%
\bibitem [{\citenamefont {Kaminski}\ \emph {et~al.}(2012)\citenamefont
  {Kaminski}, \citenamefont {Kampel}, \citenamefont {Steenstrup}, \citenamefont
  {Griesmaier}, \citenamefont {Polzik},\ and\ \citenamefont
  {M{\"{u}}ller}}]{Kaminski2012}%
  \BibitemOpen
  \bibfield  {author} {\bibinfo {author} {\bibfnamefont {F.}~\bibnamefont
  {Kaminski}}, \bibinfo {author} {\bibfnamefont {N.~S.}\ \bibnamefont
  {Kampel}}, \bibinfo {author} {\bibfnamefont {M.~P.~H.}\ \bibnamefont
  {Steenstrup}}, \bibinfo {author} {\bibfnamefont {A.}~\bibnamefont
  {Griesmaier}}, \bibinfo {author} {\bibfnamefont {E.~S.}\ \bibnamefont
  {Polzik}},\ and\ \bibinfo {author} {\bibfnamefont {J.~H.}\ \bibnamefont
  {M{\"{u}}ller}},\ }\href {https://doi.org/10.1140/epjd/e2012-30038-0}
  {\bibfield  {journal} {\bibinfo  {journal} {The European Physical Journal D}\
  }\textbf {\bibinfo {volume} {66}},\ \bibinfo {pages} {227} (\bibinfo {year}
  {2012})}\BibitemShut {NoStop}%
\bibitem [{\citenamefont {Eto}\ \emph {et~al.}(2019)\citenamefont {Eto},
  \citenamefont {Shibayama}, \citenamefont {Shibata}, \citenamefont {Torii},
  \citenamefont {Nabeta}, \citenamefont {Saito},\ and\ \citenamefont
  {Hirano}}]{Eto2019}%
  \BibitemOpen
  \bibfield  {author} {\bibinfo {author} {\bibfnamefont {Y.}~\bibnamefont
  {Eto}}, \bibinfo {author} {\bibfnamefont {H.}~\bibnamefont {Shibayama}},
  \bibinfo {author} {\bibfnamefont {K.}~\bibnamefont {Shibata}}, \bibinfo
  {author} {\bibfnamefont {A.}~\bibnamefont {Torii}}, \bibinfo {author}
  {\bibfnamefont {K.}~\bibnamefont {Nabeta}}, \bibinfo {author} {\bibfnamefont
  {H.}~\bibnamefont {Saito}},\ and\ \bibinfo {author} {\bibfnamefont
  {T.}~\bibnamefont {Hirano}},\ }\href
  {https://doi.org/10.1103/PhysRevLett.122.245301} {\bibfield  {journal}
  {\bibinfo  {journal} {Physical Review Letters}\ }\textbf {\bibinfo {volume}
  {122}},\ \bibinfo {pages} {245301} (\bibinfo {year} {2019})}\BibitemShut
  {NoStop}%
\bibitem [{\citenamefont {Burnett}\ \emph {et~al.}(1996)\citenamefont
  {Burnett}, \citenamefont {Julienne},\ and\ \citenamefont
  {Suominen}}]{Burnett1996}%
  \BibitemOpen
  \bibfield  {author} {\bibinfo {author} {\bibfnamefont {K.}~\bibnamefont
  {Burnett}}, \bibinfo {author} {\bibfnamefont {P.~S.}\ \bibnamefont
  {Julienne}},\ and\ \bibinfo {author} {\bibfnamefont {K.-A.}\ \bibnamefont
  {Suominen}},\ }\href {https://doi.org/10.1103/PhysRevLett.77.1416} {\bibfield
   {journal} {\bibinfo  {journal} {Physical Review Letters}\ }\textbf {\bibinfo
  {volume} {77}},\ \bibinfo {pages} {1416} (\bibinfo {year}
  {1996})}\BibitemShut {NoStop}%
\bibitem [{\citenamefont {Fuhrmanek}\ \emph {et~al.}(2012)\citenamefont
  {Fuhrmanek}, \citenamefont {Bourgain}, \citenamefont {Sortais},\ and\
  \citenamefont {Browaeys}}]{Fuhrmanek2012}%
  \BibitemOpen
  \bibfield  {author} {\bibinfo {author} {\bibfnamefont {A.}~\bibnamefont
  {Fuhrmanek}}, \bibinfo {author} {\bibfnamefont {R.}~\bibnamefont {Bourgain}},
  \bibinfo {author} {\bibfnamefont {Y.~R.~P.}\ \bibnamefont {Sortais}},\ and\
  \bibinfo {author} {\bibfnamefont {A.}~\bibnamefont {Browaeys}},\ }\href
  {https://doi.org/10.1103/PhysRevA.85.062708} {\bibfield  {journal} {\bibinfo
  {journal} {Physical Review A}\ }\textbf {\bibinfo {volume} {85}},\ \bibinfo
  {pages} {062708} (\bibinfo {year} {2012})}\BibitemShut {NoStop}%
\bibitem [{\citenamefont {Geremia}\ \emph {et~al.}(2006)\citenamefont
  {Geremia}, \citenamefont {Stockton},\ and\ \citenamefont
  {Mabuchi}}]{Geremia2006}%
  \BibitemOpen
  \bibfield  {author} {\bibinfo {author} {\bibfnamefont {J.~M.}\ \bibnamefont
  {Geremia}}, \bibinfo {author} {\bibfnamefont {J.~K.}\ \bibnamefont
  {Stockton}},\ and\ \bibinfo {author} {\bibfnamefont {H.}~\bibnamefont
  {Mabuchi}},\ }\href {https://doi.org/10.1103/PhysRevA.73.042112} {\bibfield
  {journal} {\bibinfo  {journal} {Physical Review A}\ }\textbf {\bibinfo
  {volume} {73}},\ \bibinfo {pages} {042112} (\bibinfo {year}
  {2006})}\BibitemShut {NoStop}%
\bibitem [{\citenamefont {Jammi}\ \emph {et~al.}(2018)\citenamefont {Jammi},
  \citenamefont {Pyragius}, \citenamefont {Bason}, \citenamefont {Florez},\
  and\ \citenamefont {Fernholz}}]{Jammi2018}%
  \BibitemOpen
  \bibfield  {author} {\bibinfo {author} {\bibfnamefont {S.}~\bibnamefont
  {Jammi}}, \bibinfo {author} {\bibfnamefont {T.}~\bibnamefont {Pyragius}},
  \bibinfo {author} {\bibfnamefont {M.~G.}\ \bibnamefont {Bason}}, \bibinfo
  {author} {\bibfnamefont {H.~M.}\ \bibnamefont {Florez}},\ and\ \bibinfo
  {author} {\bibfnamefont {T.}~\bibnamefont {Fernholz}},\ }\href
  {https://doi.org/10.1103/PhysRevA.97.043416} {\bibfield  {journal} {\bibinfo
  {journal} {Physical Review A}\ }\textbf {\bibinfo {volume} {97}},\ \bibinfo
  {pages} {043416} (\bibinfo {year} {2018})}\BibitemShut {NoStop}%
\bibitem [{\citenamefont {Eto}\ \emph {et~al.}(2013)\citenamefont {Eto},
  \citenamefont {Ikeda}, \citenamefont {Suzuki}, \citenamefont {Hasegawa},
  \citenamefont {Tomiyama}, \citenamefont {Sekine}, \citenamefont {Sadgrove},\
  and\ \citenamefont {Hirano}}]{Eto2013}%
  \BibitemOpen
  \bibfield  {author} {\bibinfo {author} {\bibfnamefont {Y.}~\bibnamefont
  {Eto}}, \bibinfo {author} {\bibfnamefont {H.}~\bibnamefont {Ikeda}}, \bibinfo
  {author} {\bibfnamefont {H.}~\bibnamefont {Suzuki}}, \bibinfo {author}
  {\bibfnamefont {S.}~\bibnamefont {Hasegawa}}, \bibinfo {author}
  {\bibfnamefont {Y.}~\bibnamefont {Tomiyama}}, \bibinfo {author}
  {\bibfnamefont {S.}~\bibnamefont {Sekine}}, \bibinfo {author} {\bibfnamefont
  {M.}~\bibnamefont {Sadgrove}},\ and\ \bibinfo {author} {\bibfnamefont
  {T.}~\bibnamefont {Hirano}},\ }\href
  {https://doi.org/10.1103/PhysRevA.88.031602} {\bibfield  {journal} {\bibinfo
  {journal} {Physical Review A}\ }\textbf {\bibinfo {volume} {88}},\ \bibinfo
  {pages} {031602(R)} (\bibinfo {year} {2013})}\BibitemShut {NoStop}%
\bibitem [{\citenamefont {Awschalom}\ \emph {et~al.}(1988)\citenamefont
  {Awschalom}, \citenamefont {Rozen}, \citenamefont {Ketchen}, \citenamefont
  {Gallagher}, \citenamefont {Kleinsasser}, \citenamefont {Sandstrom},\ and\
  \citenamefont {Bumble}}]{Awschalom1988}%
  \BibitemOpen
  \bibfield  {author} {\bibinfo {author} {\bibfnamefont {D.~D.}\ \bibnamefont
  {Awschalom}}, \bibinfo {author} {\bibfnamefont {J.~R.}\ \bibnamefont
  {Rozen}}, \bibinfo {author} {\bibfnamefont {M.~B.}\ \bibnamefont {Ketchen}},
  \bibinfo {author} {\bibfnamefont {W.~J.}\ \bibnamefont {Gallagher}}, \bibinfo
  {author} {\bibfnamefont {A.~W.}\ \bibnamefont {Kleinsasser}}, \bibinfo
  {author} {\bibfnamefont {R.~L.}\ \bibnamefont {Sandstrom}},\ and\ \bibinfo
  {author} {\bibfnamefont {B.}~\bibnamefont {Bumble}},\ }\href
  {https://doi.org/10.1063/1.100291} {\bibfield  {journal} {\bibinfo  {journal}
  {Applied Physics Letters}\ }\textbf {\bibinfo {volume} {53}},\ \bibinfo
  {pages} {2108} (\bibinfo {year} {1988})}\BibitemShut {NoStop}%
\bibitem [{\citenamefont {Wakai}\ and\ \citenamefont {{Van
  Harlingen}}(1988)}]{Wakai1988}%
  \BibitemOpen
  \bibfield  {author} {\bibinfo {author} {\bibfnamefont {R.~T.}\ \bibnamefont
  {Wakai}}\ and\ \bibinfo {author} {\bibfnamefont {D.~J.}\ \bibnamefont {{Van
  Harlingen}}},\ }\href {https://doi.org/10.1063/1.99197} {\bibfield  {journal}
  {\bibinfo  {journal} {Applied Physics Letters}\ }\textbf {\bibinfo {volume}
  {52}},\ \bibinfo {pages} {1182} (\bibinfo {year} {1988})}\BibitemShut
  {NoStop}%
\bibitem [{\citenamefont {Boyd}\ \emph {et~al.}(2006)\citenamefont {Boyd},
  \citenamefont {Zelevinsky}, \citenamefont {Ludlow}, \citenamefont {Foreman},
  \citenamefont {Blatt}, \citenamefont {Ido},\ and\ \citenamefont
  {Ye}}]{Boyd2006}%
  \BibitemOpen
  \bibfield  {author} {\bibinfo {author} {\bibfnamefont {M.~M.}\ \bibnamefont
  {Boyd}}, \bibinfo {author} {\bibfnamefont {T.}~\bibnamefont {Zelevinsky}},
  \bibinfo {author} {\bibfnamefont {A.~D.}\ \bibnamefont {Ludlow}}, \bibinfo
  {author} {\bibfnamefont {S.~M.}\ \bibnamefont {Foreman}}, \bibinfo {author}
  {\bibfnamefont {S.}~\bibnamefont {Blatt}}, \bibinfo {author} {\bibfnamefont
  {T.}~\bibnamefont {Ido}},\ and\ \bibinfo {author} {\bibfnamefont
  {J.}~\bibnamefont {Ye}},\ }\href {https://doi.org/10.1126/science.1133732}
  {\bibfield  {journal} {\bibinfo  {journal} {Science}\ }\textbf {\bibinfo
  {volume} {314}},\ \bibinfo {pages} {1430} (\bibinfo {year}
  {2006})}\BibitemShut {NoStop}%
\bibitem [{\citenamefont {Eto}\ \emph {et~al.}(2014)\citenamefont {Eto},
  \citenamefont {Sadgrove}, \citenamefont {Hasegawa}, \citenamefont {Saito},\
  and\ \citenamefont {Hirano}}]{Eto2014}%
  \BibitemOpen
  \bibfield  {author} {\bibinfo {author} {\bibfnamefont {Y.}~\bibnamefont
  {Eto}}, \bibinfo {author} {\bibfnamefont {M.}~\bibnamefont {Sadgrove}},
  \bibinfo {author} {\bibfnamefont {S.}~\bibnamefont {Hasegawa}}, \bibinfo
  {author} {\bibfnamefont {H.}~\bibnamefont {Saito}},\ and\ \bibinfo {author}
  {\bibfnamefont {T.}~\bibnamefont {Hirano}},\ }\href
  {https://doi.org/10.1103/PhysRevA.90.013626} {\bibfield  {journal} {\bibinfo
  {journal} {Physical Review A}\ }\textbf {\bibinfo {volume} {90}},\ \bibinfo
  {pages} {013626} (\bibinfo {year} {2014})}\BibitemShut {NoStop}%
\bibitem [{\citenamefont {Wolf}\ \emph {et~al.}(2015)\citenamefont {Wolf},
  \citenamefont {Neumann}, \citenamefont {Nakamura}, \citenamefont {Sumiya},
  \citenamefont {Ohshima}, \citenamefont {Isoya},\ and\ \citenamefont
  {Wrachtrup}}]{Wolf2015}%
  \BibitemOpen
  \bibfield  {author} {\bibinfo {author} {\bibfnamefont {T.}~\bibnamefont
  {Wolf}}, \bibinfo {author} {\bibfnamefont {P.}~\bibnamefont {Neumann}},
  \bibinfo {author} {\bibfnamefont {K.}~\bibnamefont {Nakamura}}, \bibinfo
  {author} {\bibfnamefont {H.}~\bibnamefont {Sumiya}}, \bibinfo {author}
  {\bibfnamefont {T.}~\bibnamefont {Ohshima}}, \bibinfo {author} {\bibfnamefont
  {J.}~\bibnamefont {Isoya}},\ and\ \bibinfo {author} {\bibfnamefont
  {J.}~\bibnamefont {Wrachtrup}},\ }\href
  {https://doi.org/10.1103/PhysRevX.5.041001} {\bibfield  {journal} {\bibinfo
  {journal} {Physical Review X}\ }\textbf {\bibinfo {volume} {5}},\ \bibinfo
  {pages} {041001} (\bibinfo {year} {2015})}\BibitemShut {NoStop}%
\bibitem [{\citenamefont {M{\"{u}}ck}\ \emph {et~al.}(2001)\citenamefont
  {M{\"{u}}ck}, \citenamefont {Kycia},\ and\ \citenamefont
  {Clarke}}]{Muck2001}%
  \BibitemOpen
  \bibfield  {author} {\bibinfo {author} {\bibfnamefont {M.}~\bibnamefont
  {M{\"{u}}ck}}, \bibinfo {author} {\bibfnamefont {J.~B.}\ \bibnamefont
  {Kycia}},\ and\ \bibinfo {author} {\bibfnamefont {J.}~\bibnamefont
  {Clarke}},\ }\href {https://doi.org/10.1063/1.1347384} {\bibfield  {journal}
  {\bibinfo  {journal} {Applied Physics Letters}\ }\textbf {\bibinfo {volume}
  {78}},\ \bibinfo {pages} {967} (\bibinfo {year} {2001})}\BibitemShut
  {NoStop}%
\bibitem [{\citenamefont {Koch}\ \emph {et~al.}(1981)\citenamefont {Koch},
  \citenamefont {{Van Harlingen}},\ and\ \citenamefont {Clarke}}]{Koch1981}%
  \BibitemOpen
  \bibfield  {author} {\bibinfo {author} {\bibfnamefont {R.~H.}\ \bibnamefont
  {Koch}}, \bibinfo {author} {\bibfnamefont {D.~J.}\ \bibnamefont {{Van
  Harlingen}}},\ and\ \bibinfo {author} {\bibfnamefont {J.}~\bibnamefont
  {Clarke}},\ }\href {https://doi.org/10.1063/1.92345} {\bibfield  {journal}
  {\bibinfo  {journal} {Applied Physics Letters}\ }\textbf {\bibinfo {volume}
  {38}},\ \bibinfo {pages} {380} (\bibinfo {year} {1981})}\BibitemShut
  {NoStop}%
\bibitem [{\citenamefont {Koch}\ \emph {et~al.}(1980)\citenamefont {Koch},
  \citenamefont {{Van Harlingen}},\ and\ \citenamefont {Clarke}}]{Koch1980}%
  \BibitemOpen
  \bibfield  {author} {\bibinfo {author} {\bibfnamefont {R.~H.}\ \bibnamefont
  {Koch}}, \bibinfo {author} {\bibfnamefont {D.~J.}\ \bibnamefont {{Van
  Harlingen}}},\ and\ \bibinfo {author} {\bibfnamefont {J.}~\bibnamefont
  {Clarke}},\ }\href {https://doi.org/10.1103/PhysRevLett.45.2132} {\bibfield
  {journal} {\bibinfo  {journal} {Physical Review Letters}\ }\textbf {\bibinfo
  {volume} {45}},\ \bibinfo {pages} {2132} (\bibinfo {year}
  {1980})}\BibitemShut {NoStop}%
\bibitem [{\citenamefont {Puentes}\ \emph {et~al.}(2013)\citenamefont
  {Puentes}, \citenamefont {Colangelo}, \citenamefont {Sewell},\ and\
  \citenamefont {Mitchell}}]{Puentes2013}%
  \BibitemOpen
  \bibfield  {author} {\bibinfo {author} {\bibfnamefont {G.}~\bibnamefont
  {Puentes}}, \bibinfo {author} {\bibfnamefont {G.}~\bibnamefont {Colangelo}},
  \bibinfo {author} {\bibfnamefont {R.~J.}\ \bibnamefont {Sewell}},\ and\
  \bibinfo {author} {\bibfnamefont {M.~W.}\ \bibnamefont {Mitchell}},\ }\href
  {https://doi.org/10.1088/1367-2630/15/10/103031} {\bibfield  {journal}
  {\bibinfo  {journal} {New Journal of Physics}\ }\textbf {\bibinfo {volume}
  {15}},\ \bibinfo {pages} {103031} (\bibinfo {year} {2013})}\BibitemShut
  {NoStop}%
\end{thebibliography}
\end{document}